# Vibrations of a one-dimensional host-guest system

A.C. Sparavigna

Dipartimento di Fisica, Politecnico di Torino
C.so Duca degli Abruzzi 24, 10129 Torino, Italy

A simple model shows how it is possible to create a gap in the vibrational spectrum of a one-dimensional lattice. The proposed model is a host-guest chain having, instead of point-like masses connected by spring, massive cages hosting particles inside. We imagine the cage as a rigid box containing a mass linked by a spring to the box inner wall. The presence of guests creates an energy gap in the dispersion of vibrational frequencies. The gap is about the internal resonance of the mass hidden in the cage.

In spite of their simplicity, the study of models with masses or rods connected by springs can be quite helpful in understanding the properties of metamaterials as of other nano-engineered structures. The term "metamaterial", credited to Rodger Walser, indicates a material, which gains its properties from its structure, rather than from the properties of components [1,2]. In fact, the term "metamaterial" is commonly used for composites, which are distinguishing themselves for unusual properties. There are several examples regarding electromagnetism and elastic properties. We have, for instance, the left-handed materials possessing negative refractive index, able to affect in an uncommon way the passage of electromagnetic waves near them. [3-6]. In the case of elastic materials, a property considered as unusual is a negative Poisson's ratio. Materials with negative Poisson coefficient are named auxetics [7,12]. Among them, natural auxetics occur in biological systems too.

Metamaterials usually share similar behaviours with photonic and phononic crystals [13,14]. For instance, some metamaterials have been prepared, which are able to act as total wave reflector within certain sonic frequency ranges. These sonic materials, which are then behaving as phononic crystals, are mainly fabricated including in a hosting component some localized resonant structures [15].

As in the case of electromagnetic metamaterials, we can prepare some composites displaying an effective "negative" elastic constant, analogous to the negative refractive index [16,17]. In [16], the author is discussing the case of metamaterials, which are guest-host systems, having units possessing hidden resonant masses inside. Figures in Ref.16 are quite stimulating to study and discuss the vibrational properties of such structures. Among the many models composed of rigid cages with moving particles inside, let us use the simplest one we can imagine, that is a one-dimensional chain composed of rigid host-guest units.

Figure 1 shows the model. It is a simple spring model describing an interacting system of host cages of mass $M$ interconnected by springs having constant $K$ and guest atoms of mass $m$ attached to the cage inner walls by means of springs with constant $K'$. The one-dimensional model we consider has then rigid units and spring connections, with distance $L$ between cages. The unit cell of the lattice has a position given by the lattice indices $i-1, i, i+1, i+2,...$ If the cage is imagined as a closed box, a mass can be hidden in it. Its presence is revealed by the frequencies of the system.

Let us define $\omega_o = \sqrt{K/M}$, $\omega' = \sqrt{K'/m}$ which are the natural angular frequencies of cage and hidden masses. In the following we will use the dimensionless ratios: $k' = K'/K$, $m' = m/M$, $\Omega = \omega/\omega_o$ and $\Omega' = \omega'/\omega_o$. Let us investigate the harmonic vibrations of the chain supposed to be infinite with displacements of masses in longitudinal direction. $x_{b,i}$ is the displacement from equilibrium of one of two masses, that are the cage and the hidden mass: $b$ can have two possible determinations $M$ and $m$ for the reticular position $i$ of equilibrium.

In the case of small displacements, equations are:

$$M \ddot{x}_{M,i} + 2K x_{M,i} - K x_{M,i-1} - K x_{M,i+1} - K'(x_{m,i} - x_{M,i}) = 0$$
$$m \ddot{x}_{m,i} + K'(x_{m,i} - x_{M,i}) = 0 \qquad (1)$$

that is, using $\eta_i = x_{M,i}, \xi_i = x_{M,i} - x_{m,i}$, we have:

$$M \ddot{\eta}_i + 2K \eta_i - K \eta_{i-1} - K \eta_{i+1} - K' \xi_i = 0$$
$$m \ddot{\xi}_i + m \ddot{\eta}_i + K' \xi_i = 0 \qquad (2)$$

If we are looking for Bloch waves with wavevector $q$, it is possible to write for each lattice site:

$$\eta_i = A \exp(i\omega t - iqx) \quad ; \quad \xi_i = B \exp(i\omega t - iqx) \qquad (3)$$

and then the dispersion relations for frequency $\omega$ can be easily obtained from the dynamical equations (2). Let us consider for plotting, the reduced frequencies $\Omega' = \omega'/\omega_o$, $\Omega = \omega/\omega_o$. Dispersion relations of the chain as a function of the wavenumber $q$ are shown in Figure 2. Note the existence of a phononic gap between the two branches. This gap is about the natural frequency of the mass inside the cage. The figure is obtained assuming $m'=1/4$, $k'=1/4$. The horizontal line represents $\Omega'$, the reduced natural frequency of the hidden oscillator. At the edge of the Brillouin Zone, we have a frequency of the system almost corresponding to that of the natural oscillation $\Omega'$.

Figure 3 shows the dispersion of vibrational frequencies in several conditions. In the upper part of the figure, we see the behaviour of dispersion for three fixed values of $k'$. In each panel, $m'$ is changing. It is possible to observe that the gap increases and that the acoustic branch has a long wavelength limit possessing a sound speed decreasing with the increase of the hidden mass. In the lower part of the image, it is the value of $m'$ to be fixed and each panel shows the dispersions as the constant of the spring $k'$ is varying.

The model shows, from a macroscopic point of view, that a structure with a cage hosting a mass displays a gap in the allowed frequencies. It is therefore illustrating how phononic crystals are created by means of host-guest systems. Moreover, we have seen that an increase of the hidden mass reduces the speed of the acoustic long wavelengths, an interesting result for engineering materials with very low thermal conductivity and for the development of more efficient thermoelectric devices.

In fact, a low thermal conductivity is required for the thermoelectric conversion in solid-state heat engines. In these devices, the electron gas serves as the working fluid, converting the heat flow in electric power [18]. For thermoelectric applications, materials must have a high figure-of-merit, which is a goodness factor including the Seebeck coefficient and the electrical and thermal conductivities. A decrease of thermal conductivity means an increase of the figure-of-merit. In the case of crystalline materials, it is enough to disturb the phonon paths by disorder or lattice defects [19,20] to have a low conductivity. Unfortunately, defects decreased the charge transport too.

Therefore, the figure-of-merit can only be moderately improved by reducing the lattice thermal conductivity: to have a significantly larger goodness parameter it is necessary to improve the electrical properties [21]. The aim of recent researches is to employ the Phonon Glasses - Electronic Crystals, PGECs, where the lattice is disordered and then phonons are strongly scattered, but the electrons remain free to move. To create such structures, a possibility is the use of materials containing weakly bound atoms, "rattling" within an atomic cage. These materials have a low thermal conductivity, as that displayed by glasses, but have an electric conductivity as high as in crystals [22]. Typical of these materials are the filled skutterudites [23] and the clathrates [24], which are host-guest systems at the atomic, microscopic scale.

In host-guest lattices, the guest entities are supposed to have oscillations, so-called rattler modes, which scatter the acoustic phonons and reduce the thermal conductivity. In a resonant scattering model [25], it was hypothesized an "avoided crossing" between acoustic phonons and localized guest modes, that has a consequence of a mixing of guest and host modes with an energy exchange as a consequence. Avoided crossing was found in hydrates [26,27] and recently in a PGEC material. In Ref.28, the phonon dispersion relations of $Ba_8Ga_{16}Ge_{30}$ are showing unambiguously the theoretically predicted avoided crossing of the rattler modes and the acoustic-phonon branches. $Ba_8Ga_{16}Ge_{30}$ is a clathrate type-I structure with a host cage framework of Ga and Ge atoms holding Ba guest atoms inside the cages. The phenomenon referred as the "avoided crossing", is the same as that we show in Fig.2 and 3, that is the presence of a gap separating the two branches of frequency dispersion. This gap is created about the natural frequency of the guest. The proposed simple model helps then understanding the behaviour of some phononic lattices at a macroscopic scale and the properties of materials with rattling modes at a microscopic scale.

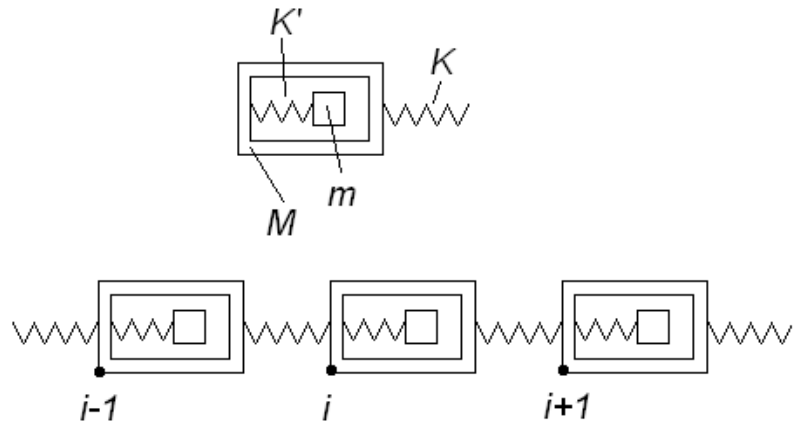

Fig.1:.The unit of the chain is composed by a cage with a hidden mass inside.

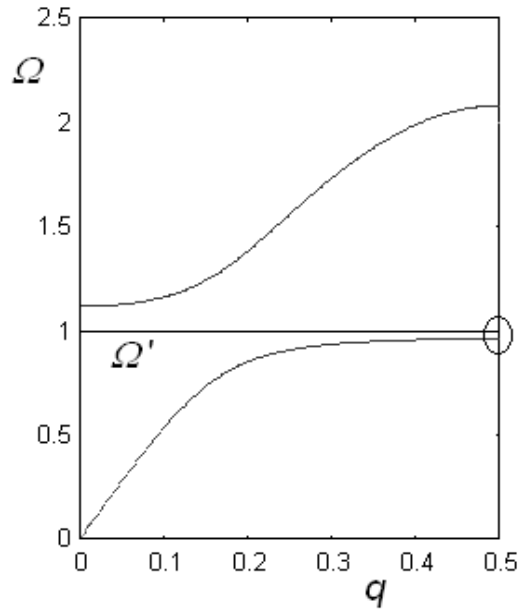

Fig.2: Dispersion of vibrational frequencies for model in Fig.1. $\Omega'$ is the natural frequency of the hidden oscillator. Note that at the edge of the Brillouin Zone (encircled) we have a frequency almost corresponding to that of the natural oscillation of the hidden mass. This means that the propagation of waves with a frequency equal to that of the internal resonance is not allowed.

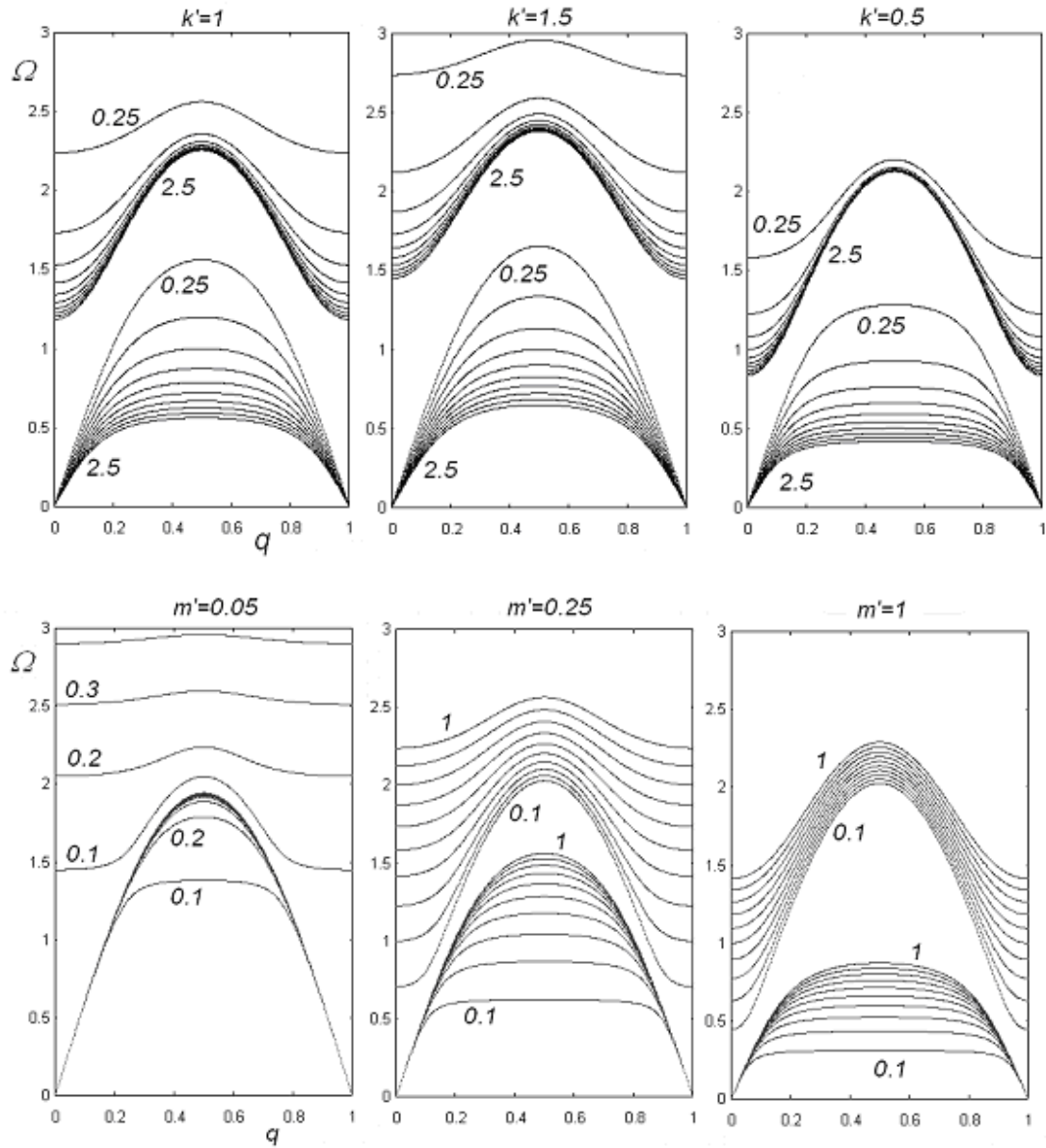

Fig.3: In the upper part of the figure, it is shown the behaviour of dispersion for three fixed values of *k'*. In each panel, *m'* is changing. It is possible to see that the gap increases and that the acoustic branch has a long wavelength limit possessing a sound speed decreasing with the increase of the hidden mass. In the lower part of the image, it is the value of *m'* to be fixed and each panel shows the dispersions as the constant of the spring *k'* is varying.